# Study of conduction, block and reflection at the excitable tissues boundary in terms of the interval model of action potential

**Sergii Kovalchuk**

**Geolab, Odessa, Ukraine**

Corresponding Author:

**telluris@meta.ua**

**Keywords:** action potential propagation, cells interaction**,** block, reflection, axiomatic model

**Abstract**

Some mechanisms of cardiac arrhythmias can be presented as a composition of elementary acts of block and reflection on the contacts of homogeneous areas of the conducting tissue. For study this phenomena we use an axiomatic one-dimensional model of interaction of cells of excitable tissue. The model has four functional parameters that determine the functional states durations of the cell.. We show that the cells of a homogeneous excitable tissue, depending on the ratio of the durations of the functional intervals, can operate in the mode of solitary wave's conduction or in one of three modes of self-generation. It is proved that the propagation of a solitary wave through the boundary of homogeneous conducting tissues can be accompanied by a block or multiplex reflection. Block and reflection are unidirectional phenomena, and there are not compatible on the same boundary. Systematized rules of transmitting, block and reflection waves at the boundary of homogeneous conducting tissues open up new possibilities for design mechanisms of generation and analyzing complex heart rate patterns.

## 1 Introduction

Usually, the basic mechanisms of arrhythmias [1] are presented as composition of elementary phenomena: unidirectional conduction block and reflection of action potential (AP) signal rhythm transformation such as Wenckebach's periodicity. In order to analyze and systematize these phenomena, we have developed the 1-D axiomatic model, in which the evolution of the action potential is represented in the form of 4 functional intervals. In the first section, we analyze the use of axiomatic models in the problems of propagation of the AP in cardiac tissue. On this basis, the requirements for functionality and the choice of parameters of the new model are built.

In one-parameter axiomatic model of Wiener–Rosenblueth's [2] the transmission of excitation in the cardiac tissue is represented as the propagation of refractoriness waves. Follow the ideas this model, Rosenblueth [3] suggested that the Wenckebach periodicity (the loss of individual beats in the periodic rhythm) as the interaction result between the refractory period of the heart tissue and stimulation rate. In the paper [4], to explain the "stop" of the exciting signal propagation, in addition to the refractoriness parameter, the concept of the duration of the active state of an excited cell is used. ~~A~~ sufficient duration of the active state allows to postpone the excitation of adjacent tissue sites until the end of the refractory state.

Experimental results [5] show that an echo signal may occur at conducting from a His bundle to the atria. It should be noted that the echo effect is controlled by drugs, which modify the parameters of AP. In [6] it has been shown experimentally that the echo signal is delayed by at least one refractory period. It has been suggested that the circulation (reverberation) of echo signals is possible in the section with the travel time exceeding the refractory period. The authors do not directly associate the formation of this signal with the reentry mechanism and suggest that retrograde propagation occurs in a different path.

The “stop” effect [4] of conduction, formation of echo signals and reverberation are formally explained by the parametric model of V. Krinsky [7],[8]. In this model, the cell can be three states: neutral (ready for activation), refractory and active. The parameters of the model are the duration of the active state interval and the duration of the refractory period. The active interval is included in the refractory period. The model operates according to the following algorithm. In order for cell 1 to excite the neighboring cell 2, cell 1 must be in an active state, and cell 2 must to be in a neutral state; that means then cell 2 must not to be in a refractory state. Since the active state has a certain duration, the excited cell 1, during its active state interval, can wait for the finish of the refractory period of the cell 2 and then initiate its.

The Wiener and Krinsky models were developed and used as a tool for the analytical study of the evolution and propagation of the action potential. Thus, it was in Wiener's model that estimates of the conditions for the formation of spiral waves were first obtained. Krinsky's model quantitatively showed the value of the ratio of the duration of the active state of the cell and the duration of the refractory period for the formation of an echo signal.

With the help of this model, it is possible to adequately explain Rosenbluth's hypothesis on the mechanism of formation of the Wenckebach period. This explanation in itself is proof that the transfer of excitation between neighboring cells can be delayed for some time until the refractory state of the excitable cell is completed.

Krinsky's axiomatic model can be given an adequate physiological meaning if the interval of high action potential values (plateau phase) to associate with the active state of cell. In this interpretation, the transfer of excitation between neighboring cells occurs under the condition of a high potential difference between these cells. When this condition is met, the cells with any allowable combination of active state duration and refractory state duration parameters can excite each other. Therefore, the boundary of two different conductive tissues is always conductive. But this means that the blockade of the conducting of AP cannot be described by the Krinsky model.

## 2 The model justification

We aim to expand Krinsky's model to analytical study the effects of conduction at the interface of inhomogeneous media. The most important of these effects is conduction block. Analysis of this effect allows us to deduce the way of expansion of the Krinsky model.

The term "conduction block” is used in a broad sense. In cardiology [9], for example, heart block means a type of heart arrhythmia. These anomalies evolve over time and depend on the cardiac rhythm. Following [9], we associate their appearances with the mechanisms of rhythm transformations. To ignore the various effects associated with the excitation rate of cells, in this paper doesn’t consider the occurrence of induced blocks [10], [11].

In our work, we assume that the block is the absence of conduction a single AP impulsion through the boundary of conducting tissues. This means that under conditions of block, cells at the contact of two different conductive tissues are unable to excite each other at least in one conductions direction. That is, conduction block is an effect that occurs in heterogeneous tissue, in the simplest case, at the boundary of homogeneous tissues.

Many studies show the relationship between an inhibition of conduction rate and the occurrence of block [12]. In particular, a computational experiment [13] shows that the propagation rate of the excitation wave depends on the uncoupling of gap junctions and the opening speed of fast ion channels in the initial phase of the AP evolution. On the other hand, there is an important observation [14] that the uncoupling of gap junctions, which causes inhibition of AP propagation, associated with an increase of growth rate of AP during the phase 0, can lead may result to unidirectional conduction's block. This implies an important conclusion for the construction of our model that a block is not a function the velocity of the AP propagation, but a block could be accompanied by the decrease of AP propagation's velocity.

In the physiological ion models [15] of AP evolution, two stages are clearly distinguished: external stimulation and autonomous forming of AP. The idea to separate the initial phase of AP evolution into slow and fast stages was proposed in the "integrate-and-fire" model [16] long before the appearance of ionic models.

In ionic models, the role of a stimulus performs current flowing through the gap junction [13]. The magnitude of this current is proportional to the potential difference of interacting cells and is inversely proportional to the ohmic resistance of the gap junction. The external stimulating current carries a positive charge, thereby performing partial depolarization of the cardiomyocyte membrane. At the moment when the transferred charge reaches the threshold level, fast ion channels open and the process of autonomous formation of the action potential of the cell begins. The charge transfer process takes a certain time and this time must be provided by the duration of the stimulating current. It is easy to see that this statement is a consequence of the well-known principle witch expressing the relation between the strength and duration of a just threshold stimulus ("strength-duration curve") [17]. In different types of cardiomyocytes, the difference in potentials between the state of depolarization and polarization ("strength") between different pairs of cells of the conducting system is insignificant. Therefore, as a first approximation, one should expect that the main factor in the possibility of excitation of neighboring cells is the excitation time factor ("duration").

We discussed the possibility of separate the process of AP growth at the starting of AP evolution into the stage of external stimulation and autonomous formation of AP, using the example of the ion model, in which the cell was stimulated by a current by gap junctions. According to field theory [18], [19], the transfer of excitation between cells is carried out by an electric field, partially depolarizing the membrane. The time of the external initialization stage, in this case, is determined by the potential difference between the cells and the membrane capacity.

Thus, within the framework of various physical assumptions, the required duration of external influence (under normal conditions of cell interaction) can serve as a parameter that, in the first approximation, characterizes the condition of cell excitation. In these terms, the excitation of cell 2 from the neighboring cell 1 cannot be carried out if the duration of the active state of cell 1 is less than the duration of external influence required for cell 2. Otherwise, a conduction block occurs on the contact of cells. So, the block is a function of the duration of external initialization and the duration of the active state of the cell.

The time for equalizing the potentials of interacting cells is equal to the switching time of excitation, which determines the propagation velocity of the AP wave. The threshold potential for switching on fast channels is much lower (more negative) than the potential of the active state of the cell. Therefore, the duration of the interval of the required external influence is less than the total time for equalizing the potentials of the interacting cells. Thus, we found that in order to build a model of AP that allows for the existence of a conduction block, it is necessary to take into account the duration of the external influence and the process of switching excitation of the cells as different factors. Within the framework of axiomatic models, taking these factors into account can be presented as parameters characterizing the duration of the corresponding states of the cell.

A mathematical model [20] close to the context of our analysis is offered in the form of a cyclic hybrid automata (HA). In accordance with three threshold levels, the evolution of AP is represented by sequential switching of 4 modes, which the authors refer to as stimulated (ST), upstroke (UP), early repolarization and plateau (EP), final repolarization and resting (FR). In each mode, the change in AP is determined by formal equations that simulate the main transmembrane ion currents. The parameters of the model are the values of the threshold levels of AP and the coefficients of the equations. Known examples [21] use the HA model in algorithms for numerical simulation of conduction in complex systems, which include the interaction of several organs.

In accordance with our reasoning, the initial part of growth in the AP shape of this model is divided into two stages: stimulated (ST) and upstroke (UP). The first stage of stimulation (ST) is decisive. The success of its completion determines the ability to implement the upstroke (UP) mode and other subsequent modes. Therefore, in the HA model, a situation is possible when there is no conduction between cells (due to insufficient stimulation).

## 3 Axiomatic interval model of interaction of cells of excitable tissue

### 3.1 The model definition

In the proposed model, the conditions for the interaction of neighboring cells are determined by low (***thr-L***) and high (***thr-H***) threshold levels of AP. The two threshold levels divide the AP shape into five parts. From this division we compose 4 intervals $(f, u, h, x)$, durations of them are the parameters of the model. In fig. 1 shows the correspondence allocated intervals with the elements of the AP shapes. The interval $f$ external initialization of the AP continues from the beginning of the AP until the low threshold ***thL*** is reached. The interval $u$ of AP fast growth ends when the potential reaches a high threshold level ***thH***. The interval $h$ of the high (positive) potential finished when the AP turns out to be below the level of the high threshold ***thH***. The relaxation interval $x$ completes the AP evolution. The total duration of the intervals $(f, u, h, x)$ is equal to the duration the refractory period $R$ of the cell:

$$R = f + u + h + x \qquad (1)$$

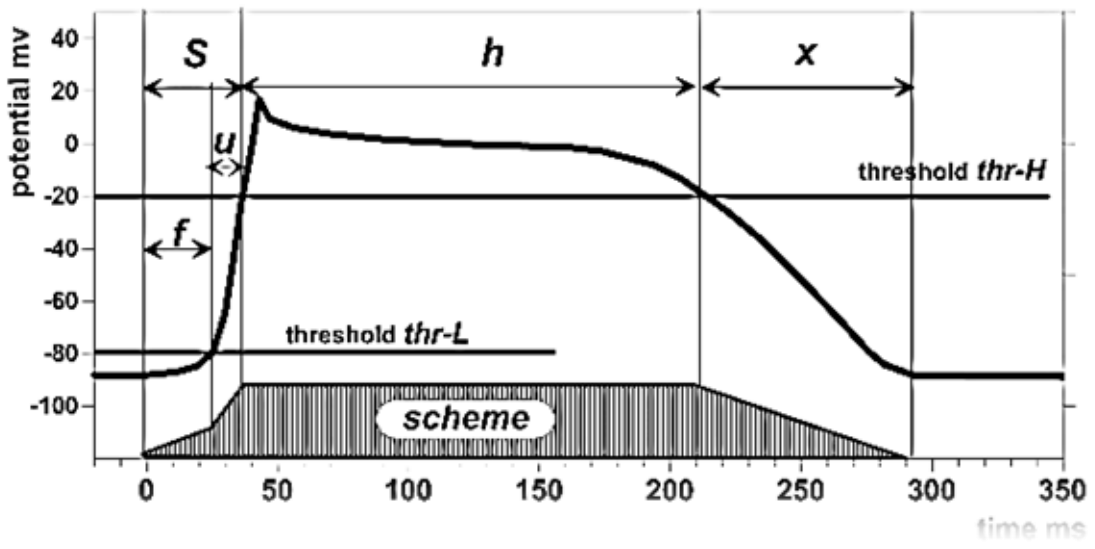

**Fig. 1** Functional intervals of Action Potential

Note that the low threshold ***thL*** and the high threshold ***thH*** formally are not the model parameters. However, the division of the entire cycle of AP evolution into intervals in accordance with some implied threshold levels means the dependence of AP changes at the level of this potential. Therefore, the durations of these intervals, in general, should be interpreted as functionals of the AP shape. On this basis, the sequence of the selected intervals can be considered the result of the functioning of the dynamic system. The highlighted intervals correspond to various functional states of the cell. In the course of its evolution, the cell and its AP successively pass through all the listed states. The evolution of AP is determined by the condition of successful completion of the external initialization interval $f$ . If this interval is successfully completed, then the further evolution of the AP occurs autonomously, i.e. the following states are sequentially switched on regardless of external conditions. In the case of unsuccessful completion of the external initialization interval $f$, the further evolution of AP immediately ends with a switching to a neutral state. Thus, the switching to the future state is determined, in the general case, by the condition of the completion of the current state of the cell.

Let's designate the transfer of excitation from cell 1 to cell 2 as mnemonic scheme (1→2). In our model, for the interaction of cells to excite (1→2) cell 2 from cell 1, three conditions must be met:

(i) cell 2 must be in a neutral state;

(ii) cell 1 must be in an active state $\boldsymbol{h}_1$ with high potential;

(iii) cell 1 must maintain high potential throughout the entire interval of external initialization of cell 2.

Fig. 2 schematically shows the interaction of two neighboring cells with different AP, which are represented by vectors of parameters $(\boldsymbol{f}_1, \boldsymbol{u}_1, \boldsymbol{h}_1, \boldsymbol{x}_1)$ and $(\boldsymbol{f}_2, \boldsymbol{u}_2, \boldsymbol{h}_2, \boldsymbol{x}_2)$. Assume that at the beginning time $\boldsymbol{t} = 0$ cell 2 is in a neutral state, and excitation is initiated in cell 1 under the action of an external stimulus. So cell 1 enters to the state $\boldsymbol{f}_1$ of external initial initialization. This state continues for the interval $\boldsymbol{f}_1$ of external initialization. After finishing of this interval, cell 1 goes through the interval $\boldsymbol{u}_1$ of AP fast growth. In the moment $\boldsymbol{t} = \boldsymbol{f}_1 + \boldsymbol{u}_1$ cell 1 enters the active state $\boldsymbol{h}_1$ of a high level of AP. According to (II), in this state, cell 1 is able to initiate the excitation of neighboring cell 2. Since cell 2 at this time is in a neutral state, the process $\boldsymbol{f}_2$ external initialization of cell 2 starts. At this moment, as shown in Fig. 2, the beginning of the interval $\boldsymbol{h}_1$ active state of cell 1 coincides with the beginning of the interval $\boldsymbol{f}_2$ external initialization of cell 2. Further, the evolution of cell2 AP obeys condition (III), according to which external initialization can be completed only if the activity state $\boldsymbol{h}_1$ of the excitatory cell1 is maintained throughout the entire interval $\boldsymbol{f}_2$. Therefore, the condition for excitation (1→2) is the inequality:

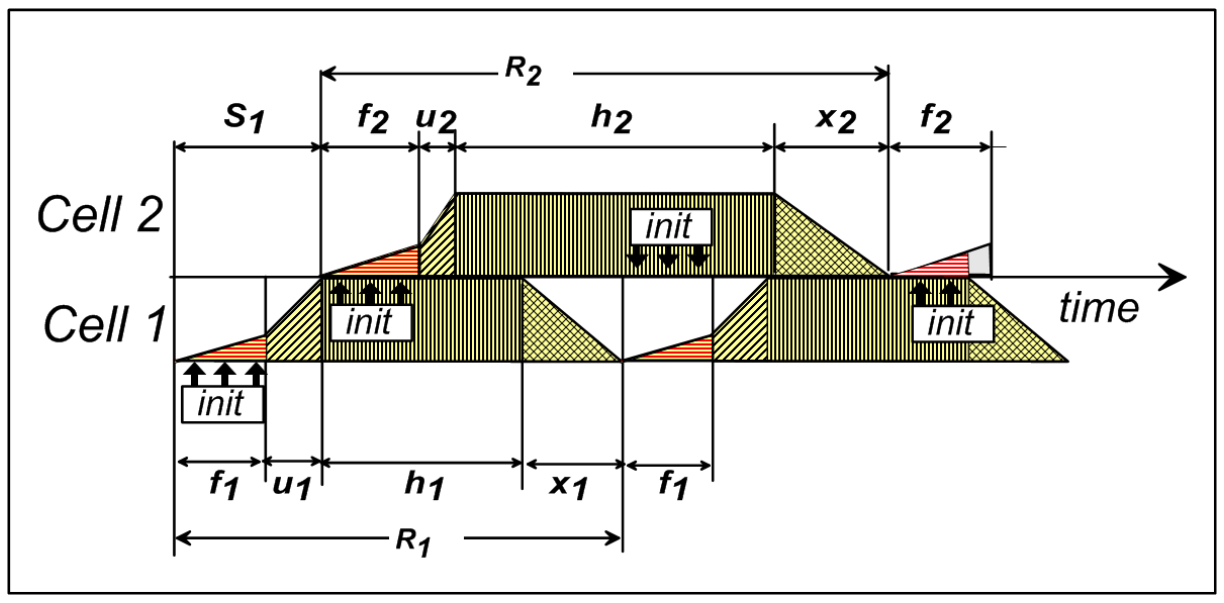

**Fig. 2** Scheme of excitation ctll1→cell2

$$\boldsymbol{h}_1 \geq \boldsymbol{f}_2 \tag{2}$$

Obviously, if we consider the excitation and conduction from cell 2 to cell 1, then its condition is the inequality:

$$\boldsymbol{h}_2 \geq \boldsymbol{f}_1 \tag{3}$$

Inequalities (2), (3) have the meaning of the necessary conditions for mutual excitation of neighboring cells, when the process of cell excitation begins at the moment the excitatory cell transitions to an active state.

When the some cells interact, it is possible that the cells re-excite according to the scheme (1→2→1). This event occurs at the moment when cell 1 leaves the refractory state and remains the required time of the active state of cell 2 for the external initialization of cell 1. As can be seen in Fig. 2, this sentence means the condition

$$(\boldsymbol{f}_1 + \boldsymbol{u}_1) + (\boldsymbol{f}_2 + \boldsymbol{u}_2) + \boldsymbol{h}_2 \geq \boldsymbol{R}_1 + \boldsymbol{f}_1 \tag{4}$$

After simple transformations, condition (4) of re-excite according to the scheme (1→2→1)) can be represented as:

$$R_2 - R_1 \geq x_2 - u_1 \qquad (5)$$

## 3.2 The modes of propagation of excitation in 1-D homogeneous media

In a homogeneous tissue, all cells have the same AP shape and, accordingly, are characterized by the same model parameters. Therefore, condition (2) of excitation (1→2) of cell 2 from the side of neighboring cell 1 for a homogeneous tissue makes sense for the condition of propagation of excitation:

$$h \geq f \qquad (6)$$

In a homogeneous tissue, the refractory periods of all cells are the same. Therefore, condition (5) of re -excitation (backward excitation) in a homogeneous tissue takes the form:

$$u \geq x \qquad (7)$$

The combination of conditions for the fulfillment of inequalities (6) and (7) determines 4 types of AP shapes, which correspond to the following modes of propagation of electric oscillations in a homogeneous conducting tissue. For purpose of studying the effects arising at the contacts of homogeneous conducting tissues, only the modes of operation of the cells between which the conduction is carried out are meaningful, i.e. inequality (6) is valid.

***(a) Mode of solitary waves.***

In this mode, the AP parameters of the tissue satisfy inequality (6) and do not satisfy inequality (7). Neighboring cells of the tissue are able to excite each other, i.e. the tissue is conductive.

A switching wave, which shape does not change during propagation, will be called a solitary wave. The constancy of the waveform propagating in the excitable tissue means the absence of self-generation, i.e. non-fulfillment of condition (7). The velocity of propagation of excitation in a one-dimensional media is determined by the delay in switching of interacting cells. As can be seen from Fig.2, in the notation of the model, this delay $s$, which it makes sense to call the start interval, is equal to the sum of the durations of the intervals and $f$, i.o. $s = f + u$. Fig. 3 illustrates the propagation and annihilation of oncoming waves. The abscissa shows the cell numbers, the ordinate shows the time. On the left, wave # 1 travels in the positive direction, wave # 2 travels in the opposite direction. At the contact between cells 7 and 8, these waves meet. Cells 7 and 8 are simultaneously in an excited state with a shift not exceeding the switching interval $s$, and cannot excite each other. Therefore, further propagation of each wave is not possible: the waves annihilate. In the AP shape of cells with solitary wave mode, a very short starting interval corresponding to a high conduction rate, a prolonged active state, and a gradual decrease in AP over a rather long relaxation interval are clearly manifested. Obviously, the propagation of solitary waves occurs in muscle tissue and atypical cardiomyocytes. It is less obvious that this group also includes tissue cardiomyocytes with relatively slow conduction, but which are characterized by a very steep rise in the avalanche growth of AP in the phase of rapid depolarization. Probably, tissues with dissociation of gap junctions can be attributed to this type.

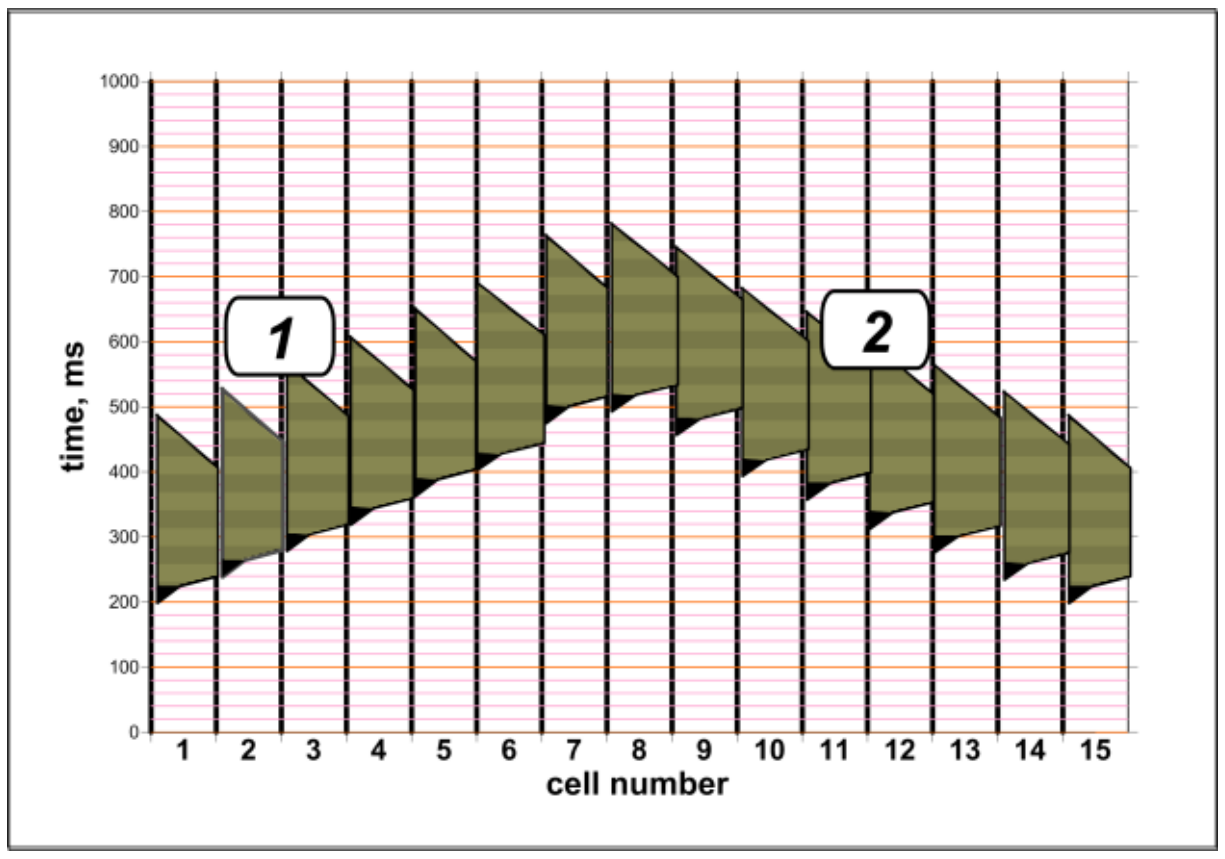

**Fig. 3.** Annihilation of oncoming waves

*(b) Auto-wave oscillation mode.*

It can be shown that the joint fulfillment of the conductivity condition (6) and the condition of reverse excitation (7) is equivalent to the condition of self-generation of oscillations. In this mode, each cell excites the neighboring cell and, after completion of its own refractory state, is excited from the previously excited cell. Thus, each pair of adjacent cells can act as a wave oscillator. The excitation wave has a leading edge followed by a periodic sequence of AP pulses. The oscillation period is equally to the refractory period of the tissue. Once arising, this wave captures all available unexcited cells. Just like in the solitary wave mode, oncoming wave packets impede further propagation of each other. As a result, clusters of oscillations with an independent wave phase appear in the wave field. We can say that such a distribution of oscillations is a one-dimensional model of cardiac fibrillation.

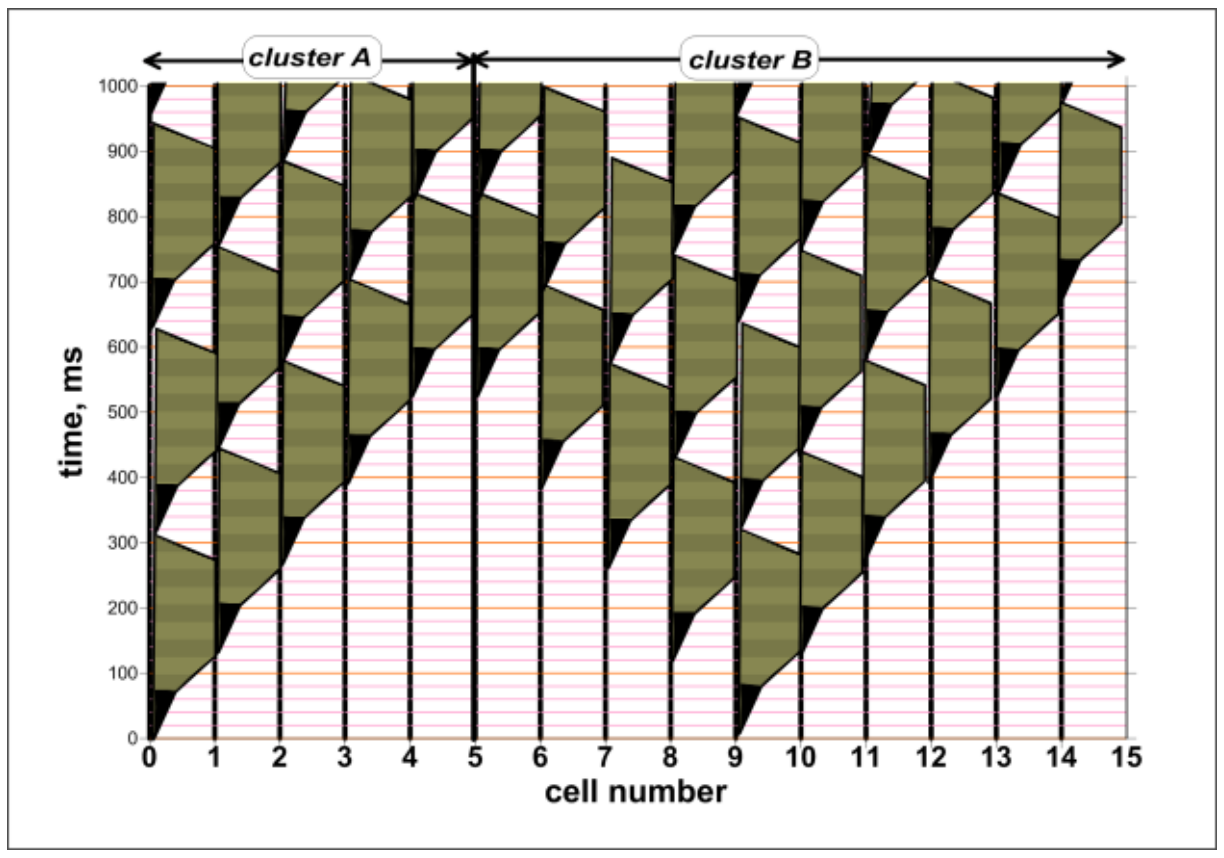

**Fig.4.** Auto-wave oscillations with two clasters formation.

Fig. 4 shows the appearance of two wave clusters. Cluster A arises as a result of the propagation of a wave packet after initial excitation in cell 1. Cluster B arises as a result of initial excitation in cell 10. At the contact of cells 5 and 6, these clusters prevent the propagation of each other. In the shape of AP of such tissue, the maximum is displaced to the left of the center. The plateau phase is relatively long. A characteristic feature is a high repolarization rate, which exceeds the rate of AP growth in phase 0 of the opening of fast channels.

# 4 Effects at the boundary of conductive tissues

## 4.1 Bidirectional transmission of the solitary waves across the boundary of conductive tissues.

In this study, we will consider only one mode (a) of propagation of the solitary AP waves in the two homogeneous conductive media, divided by the boundary. The fact that the tissue 1 and tissue 2 are conductive means that for each AP of them the condition is fulfilled like (6):

$$\boldsymbol{h}_1 \geq \boldsymbol{f}_1 \quad (8)$$

$$\boldsymbol{h}_2 \geq \boldsymbol{f}_2 \quad (9)$$

It is assumed that the wave incident from tissue 1 to the border with tissue 2. In order to carry out transmit (1→2) across the boundary, the boundary cell of the tissue 1 must excite the boundary cell of the tissue 2. As inequality (2) shows, this excitation can occur only when the duration of the AP active state interval of tissue 1 is not less than the duration of the AP external initialization interval of tissue 2. Similarly, if a wave incident from tissue 1 to the boundary with tissue 2, then the condition (2→1) of excitation of boundary cells must be satisfied. For the bidirectional transmitting through the boundary, must be met conduction conditions (2) and (3). So, in order for there to be a bidirectional transmission of solitary waves through the boundary of different conducting tissues, it is necessary to jointly fulfill inequalities (2),(3) and (8),(9). These conditions can be defined as one inequality:

$$min(h_i) \geq max(f_i) \qquad (10)$$

### 4.2 Block

Our reasoning shows that the boundary between two different conducting media does not always provide bidirectional of AP pulses transmission. Let us show that only unidirectional conduction block is possible at the boundary of the conducting tissues. So we assume that conditions of conduction (8), (9). Let wave incident from the tissue 1 to the border with tissue 2. The block of conduction (1→2) means that condition (2) for excite (1→2) is not true. That is, in the conditions of the block, the inequality inverse to inequality (2) is fulfilled:

$$f_2 > h_1$$

From this inequality and condition (8) for waves travels in tissue 1 follows that

$$f_2 \geq f_1$$

From inequality (4) (conduction of tissue 2) follows that

$$h_2 \geq f_1$$

But this inequality is the condition (3) of conduction (2→1) in the opposite direction.

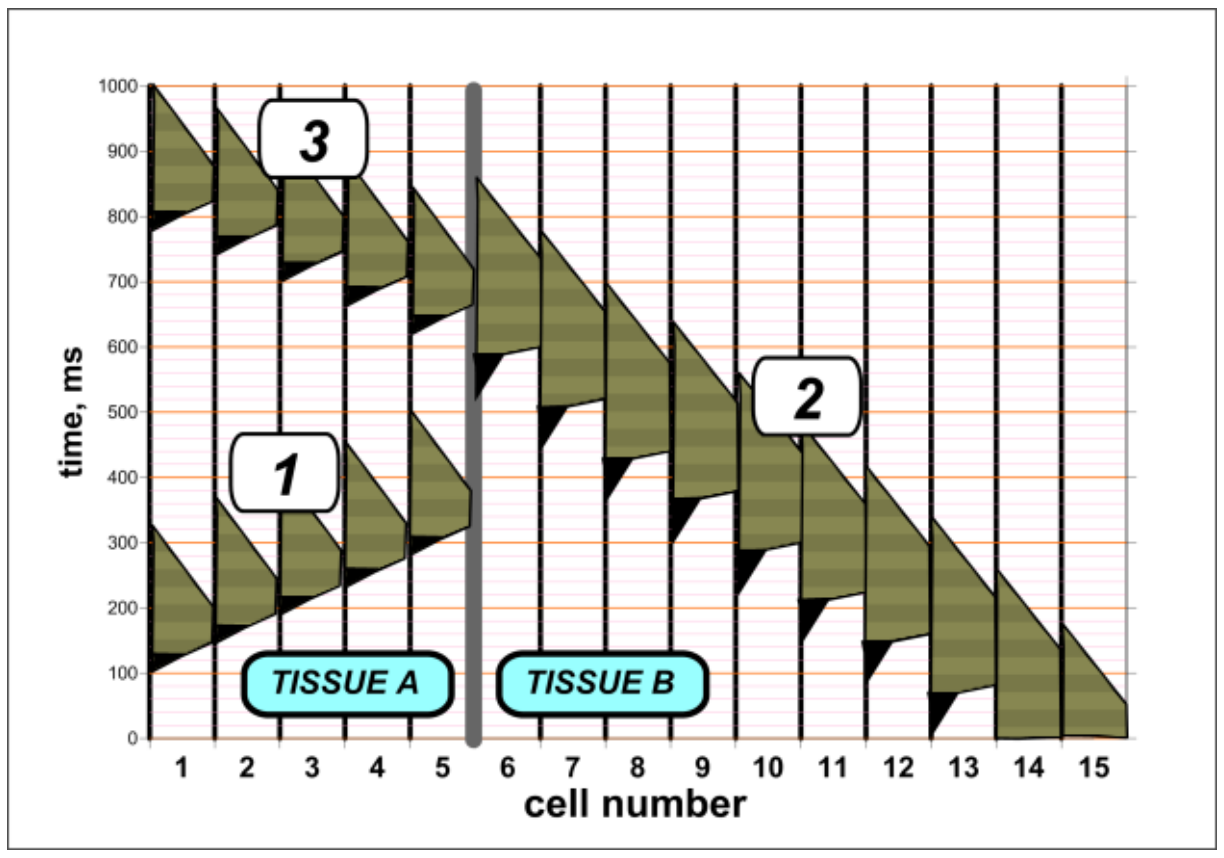

**Fig. 5.** Wave 1 blocked on the boundary of tissues A-B. Wave 2 transmitted boundary from tissue B.

Fig. 5 illustrates the formation the block of solitary wave propagation at the boundary of conducting tissues A and B, located between cells 5 and 6. The vector of AP-parameters (***f, u, h, x***) of the of tissue A is equal to (30,20,50,120) ms, the vector of parameters of tissue B is (70,10,140,120) ms. Wave 1 travel to the boundary in the tissue1 and stops propagation at the boundary. Wave 2 propagates in tissue 2 in the opposite direction without the block formation. In the next paragraph, we will show that the blockade boundary cannot be reflective in any direction.

### 4.3 Reflection

The formation of a reflected wave is possible on the boundary of conducting tissues. Let a solitary wave incident from tissue 1 to the boundary with tissue 2. As in the previous paragraph, since tissue 1 and tissue 2 are conducting, then conditions (8) and (9) are met. Since we study the solitary waves, condition of generation (7) are false for each of tissue. Let us also assume that condition (2) are true. So boundary cell 1 of tissue 1 excited the boundary cell 2 of tissue 2 and conduction (1→2) is carried out on this boundary, as a result of which a transmitted wave appears in tissue 2. As we have shown in the model definition, it is possible repeated excitation occurs in tissue 1 according to the scheme (1→2→1). We identify this process with the reflection of a wave incident from the tissue 1 to the boundary with the tissue 2. The condition for the occurrence of this event is shown in inequality

(4) or (5), which should be considered together with inequalities (2), (3) of bidirectional propagate of waves through the boundary.

Reflection of the solitary wave is always unidirectional. Let us prove this statement. Note that the general reflection condition (5) does not contain the condition for the absence of generation, i.e. conditions for the propagation of solitary waves in contacting homogeneous tissues. The conditions for the propagation of solitary waves in tissues 1 and 2 in accordance with (7) can be written in the form

$$\boldsymbol{x}_1 > \boldsymbol{u}_1$$

$$\boldsymbol{x}_2 > \boldsymbol{u}_2$$

After a series of transformations using these conditions, inequality (5) is transformed to the form

$$\boldsymbol{R_2} - \boldsymbol{R}_1 > \boldsymbol{u_2} - \boldsymbol{x_1} \tag{11}$$

Now inequality (11) expresses the condition for the formation of reflection according to the scheme (1→2→1) exclusively for solitary waves. Suppose that simultaneously with the reflection according to the scheme (1→2→1), there is also reflection in the opposite direction, which is carried out according to the scheme (2→1→2). Correspondingly changing the indices in inequality (5), we obtain that the condition for reflection (2→1→2) is the inequality

$$\boldsymbol{R_2} - \boldsymbol{R}_1 \le \boldsymbol{u_2} - \boldsymbol{x_1}$$

But this inequality is the opposite of condition (11) for the formation of reflection according to the scheme (1→2→1). The contradiction proves the proposal that the reflection of solitary waves is always unidirectional.

Reflection of solitary waves is accompanied by a delay in the formation of the reflected wave by an amount equal or more then the duration of the action potential of the incident wave. The situation is illustrated in Fig. 2. Re-excitation of the boundary cell, from which the reflected wave begins to propagate, starts after the end of its refractory period. This delay time ensures that there is no interference between the incident and reflected waves, so they do not annihilate. Reflection in the opposite direction does not occur.

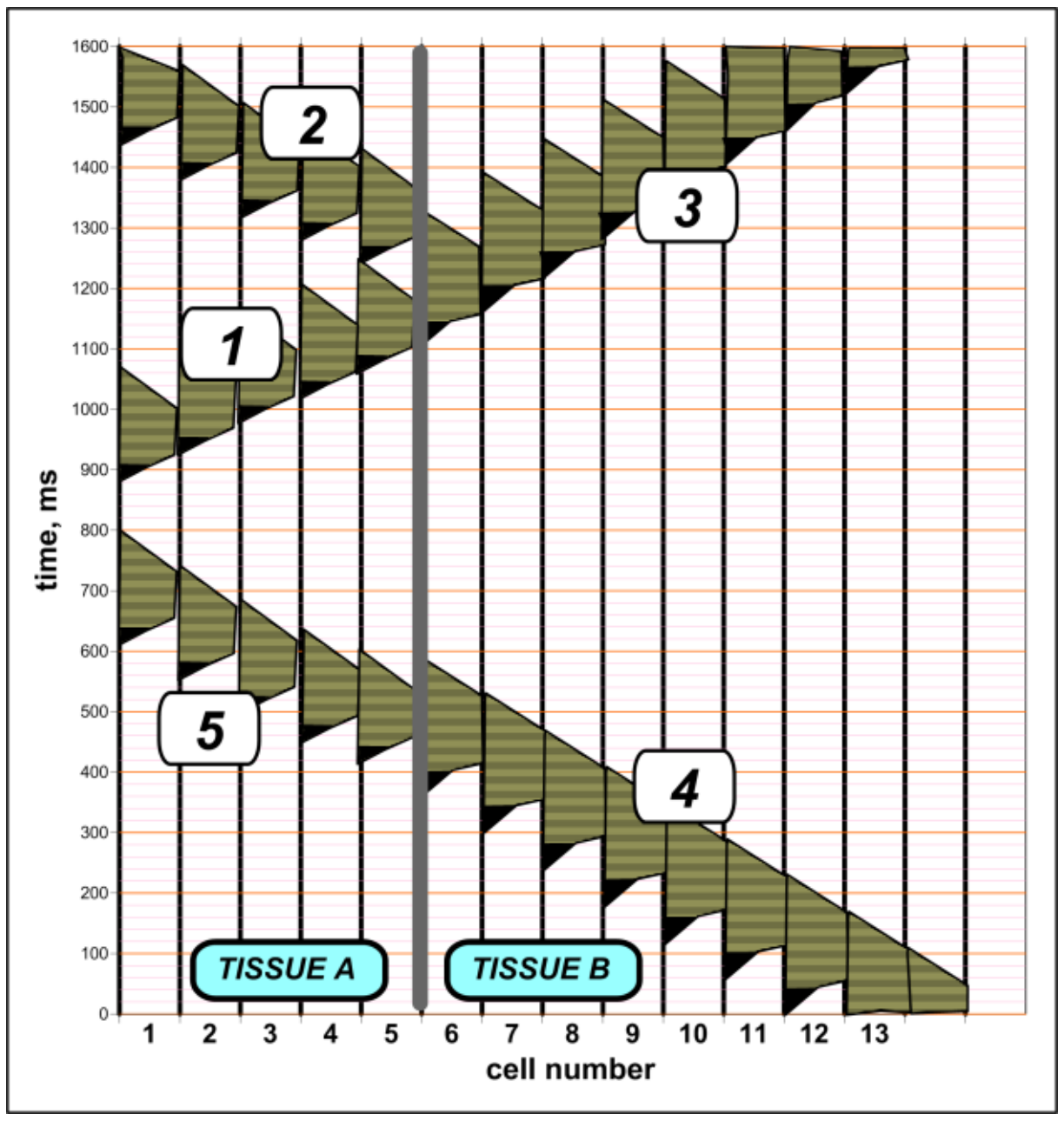

**Fig. 6.** Wave 1 incident from the tissue A to the boundary with tissue B reflected (2) and transmission (3) through the boundary. Wave 4 incident from the tissue B to the tissue A transmission (5) through the boundary without reflection.

Fig. 6 shows a situation when a wave is reflected from one side of the boundary and forms transmitted wave. When a wave incident from another side, reflection does not occur.

After determining the conditions for the formation of reflected waves, in addition to the previous paragraph, it can be argued that no reflection occurs at the boundary with the block of conduction. Suppose, for example, there is a block of conduction according to the scheme (1→2). Reflections in the forward direction according to the scheme (1→2→1) and in the opposite direction according to the scheme (2→1→2) do not occur, since both schemes include partial conduction (1→2). But this is not possible under the conditions of the blockade. It is also obvious that there can be no conduction block at the reflecting boundary.

## 4.4 Multiplex reflection

Under certain conditions, the process of border cells re-excitation can be repeated several times, causing multiple formation of a packet of transmitted and reflected waves. The condition for the first transmitted wave formation (1→2) is the condition of conductivity (2), which we will write in the form:

$$\boldsymbol{h}_1 - \boldsymbol{f}_2 \geq 0 \tag{12}$$

Suppose that condition (5) for re-excitation of cell 2 is satisfied, so, the reflected wave by the scheme (1→2→1) is arises. Starting from the moment cell 1 enters the active state, successively adding the states intervals, we obtain the condition for re-excitation of cell 2, according to the scheme (1→2→1→2)

$$\boldsymbol{h}_1 + \boldsymbol{x}_1 + \boldsymbol{f}_1 + \boldsymbol{u}_1 + \boldsymbol{h}_1 \geq \boldsymbol{R}_2 + \boldsymbol{f}_2 \tag{13}$$

This inequality can be rewritten in the form:

$$\boldsymbol{h}_1 - \boldsymbol{f}_2 \geq \boldsymbol{R}_2 - \boldsymbol{R}_1 \tag{14}$$

Generalizing (12) and (14), by induction we obtain the condition for the formation of an **n**-multiple transmitted wave:

$$\boldsymbol{h}_1 - \boldsymbol{f}_2 \geq (\boldsymbol{n}-1)(\boldsymbol{R}_2 - \boldsymbol{R}_1) \tag{15}$$

The right-hand side of (15) grows with increasing **n**, so there is such an **n** when the inequality is not satisfied. Condition (15) has the following physical interpretation. The transmitted waves impulses alternate with refractory period $\boldsymbol{R_2}$ of AP tissue 2, reflected waves - with a refractory period of AP tissue 1. Since these periods are different, the borders cells interaction on each step occurs in a different phases of AP. Therefore, there comes a moment when an attempt to re-initialize the border cell of tissue 2 is not possible due to its refractory state. Taking into account condition (5) for the appearance of the first reflection, inequality (15) can be written in a form that does not contain the duration of refractory periods:

$$\boldsymbol{h}_1 - \boldsymbol{f}_2 \geq (\boldsymbol{n}-1)(\boldsymbol{x_2} - \boldsymbol{u_1}) \tag{16}$$

Hence, we see that the multiplicity of reflection is determined by the difference in the duration of the relaxation state $\boldsymbol{x_2}$ and the duration of the AP fast growth interval $\boldsymbol{u_1}$ . In the extreme case, it may turn out that the boundary plays the role of an oscillation generator.

## 4.5 Boundary effects- summary

The results of our analytical studies of the effects arising at the boundary of homogeneous conducting tissues are combined in Fig. 7 in the form of a ray scheme. From the proposed model representation, it follows that not every boundary of conductive excitable tissues allows bilateral conduction of AP in two opposite directions (Fig.7,a). At a certain ratio of AP parameters of tissues, the propagation of the incident wave block at the boundary of these tissues. However, in this case, there is always exist counter direction of conductivity. That is, the conduction block is always unidirectional (Fig.7,b). The block boundary cannot be reflective. With some combinations of parameters, a reflection effect occurs at the tissue boundary (Fig.7,c). The reflected wave occurs with a delay equal or more then the incident wave refractory period. Therefore, a solitary reflected wave that propagates in the opposite direction does not annihilate with the incident wave. In the process of reflection, a transmitted wave necessarily appears. The reflection effect, like the block, is always unidirectional. This means that if there is a reflection from one side of the boundary, then there is no reflection in the opposite direction. A solitary incident wave can cause multiple reflections (Fig.7,d). The reflected and transmitted waves form a packet consisting of a finite number of AP pulses.

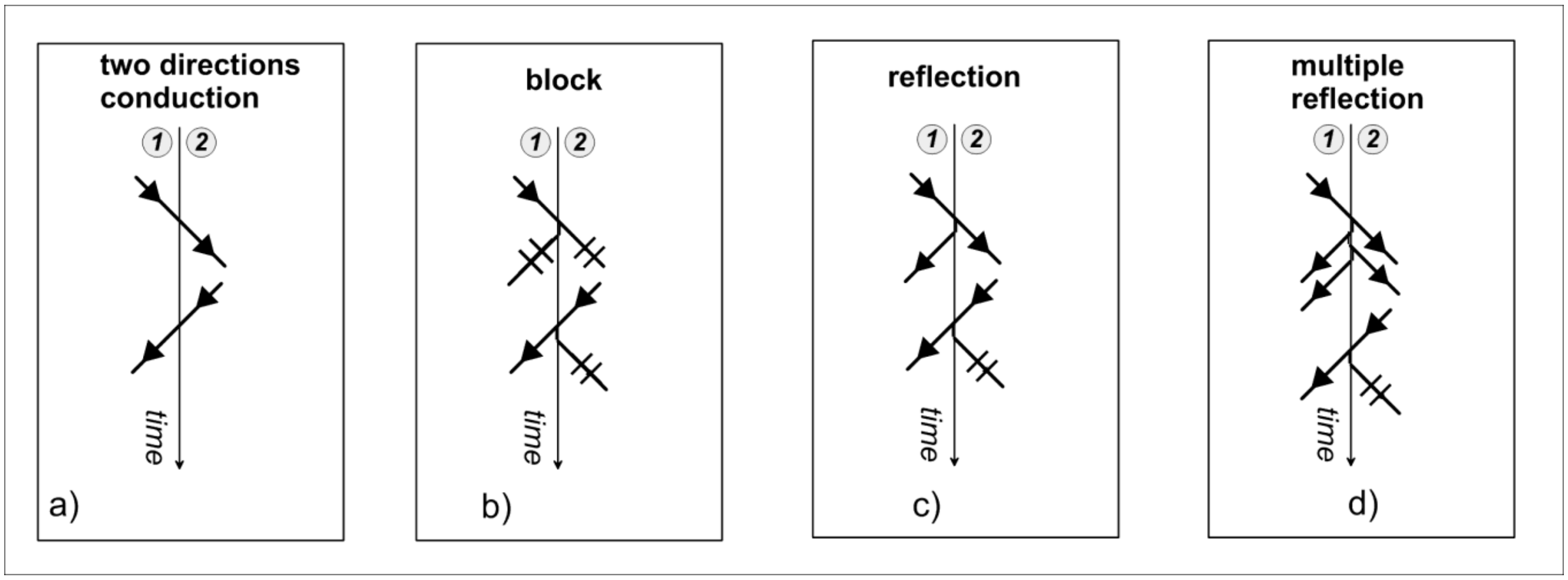

**Fig.7**. Scheme of Wave-effects on the tissues boundary**.**

# 5 Discussion

## 5.1 Modes in the media with no conduction.

When defining the model, depending on the fulfillment of the conditions for conduction (6) and the conditions of generation (7), we formally identified four modes of cells operations in a homogeneous medium. Above, we study the two modes that arise in a conductive tissues. The other two modes are formally associated with the absence of conduction in a homogeneous tissues. Our model defines the interaction of cells. Therefore, these modes cannot be explained within the framework of the model. However, the absence of interaction of cells of a homogeneous tissue does not formally prohibit the fact that a certain evolutionary process (self-oscillations) can occurs in an individual cell, which does not require external excitation to maintain. Using the analogy with the mode of auto-wave oscillations, we assume that if the inequality of carrying out (6) and the validity of the inequality of generation (7) are not fulfilled at the same time, periodic self-oscillations can arise in each cell of a homogeneous tissue. Thus, we can talk about the mode of self-oscillations of the cell.

Our assumption is supported by the following physical considerations. The rate of AP changes determines the magnitude of the charge - discharge current passing through the dielectric membrane. The proportionality of the magnitude of this current to the rate of change in potential allows us to interpret this current as a displacement current. The displacement current circuit is closed in the areas of the cell's own membrane and the membranes of neighboring cells, and is dispersed in the conducting medium surrounding the cells. In general, the conditions for the dissipation of this current are determined by the state in which the nearest cells are located. When the conduction conditions are met, the output current can maintain the active state of neighboring cells. As long as the magnitude of this current is small, it does not significantly change the evolution of the AP. When the rate of AP change in the repolarization phase reaches high values, the value of the output current increases accordingly. The conditions for its drainage into neighboring cells are difficult (there is no conduction), the current loop is closed, mainly, in the nearest areas of the cell's own membrane. This provokes the opening of fast ion channels. As a result, before the complete completion of the repolarization and refractory period, a new cell depolarization occurs. Note that repeated cell depolarization occurs without using the external initialization interval, which is used for conduction between cells. Therefore, the cell makes its own AP oscillations with a period less than the refractory period. Due to the absence of interaction, the oscillations of the cells are not synchronized. The spatial distribution of such oscillations creates an image of the phase chaos.

The shape of AP such cells is characterized by a smooth rise during depolarization and a very rapid fall in AP during the repolarization phase. The absence of conduction is manifested in the form of a shortened interval of the active state of AP. Cells operating in the mode of their own oscillations claim to be true pacemaker cells.

The last, fourth Break-mode of operation of cells in a homogeneous excitable tissue is characterized by the absence of conduction (6) and the failure to fulfill the generation condition (7). That is, the cells of the Break-mode in a homogeneous tissue do not manifest themselves in any way.

It is obvious that the cells of the considered mode of self-oscillations and the cells of the break mode, at certain ratios of the AP parameters, can interact with other cells as inclusions in other types of media, significantly changing their wave and generator properties.

### 5.2 Properties and mechanisms.

Unidirectional conduction block is a well-known phenomenon that underlies the reentry mechanism, which is used in clinical reconstructions to explain extrasystoles [1]. However, as noted in [1], extrasystoles can also be explained by the reflection of the AP wave, without using the phenomenon of conduction block. To illustrate the capabilities of the model, we will give several examples, a detailed analysis of which is beyond the scope of our publication.

Clinical interpretation of multiplex reflections leads to phenomena that resemble the description of trigger activity [1]. On the path of the excitation wave from the SA - node to the ventricle muscle tissue, there are many boundaries in the conducting system in which pathological reflection can occur.

With multiplex reflection - transmission of an excitation wave through the boundary, there are create several pulses of reflected and transmitted waves. The first impulse of the transmitted wave is, possibly, the signal of the ventricular systole. The second impulse of the transmitted wave occurs with a delay comparable to the duration of the refractory period of the heartbeat, and therefore may well be considered as a signal of extrasitole. If the event of several transmitted waves is occurrence, a packet of consecutive signals and systoles arises, which can be interpreted as a manifestation of paroxysmal tachycardia. A layer included in a homogeneous tissue can work as a waiting multivibrator. When a wave incident to the boundary of this layer, the generation of periodic oscillations arises or stops in it. This example can be considered as a one-dimensional analogue of cyclic excitation in the re-entry mechanism.

Thus, we see that the obtained rules and properties of block and reflection make it possible to reasonably design rather complex mechanisms that generate complex patterns of heart rhythm.

### 5.3 Perspectives

Within the framework of models as reaction - diffusion, the analytical study of boundary effects encounters significant mathematical difficulties. The proposed model was developed specifically for the analytical study of the boundary effects of block and reflection of the AP impulse and operates in fact with the known concepts of the electrical interaction of cells. As a result, these representations are formulated as properties of block and reflection at the boundary of homogeneous tissues. Of course, such a simplified model cannot and is not intended to explain the subtle effects of AP signal propagation in heterogeneous tissue.

The complication of the model creates prospects for further analysis, at least at a qualitative level. In particular, when constructing the model, we assumed that the shape of an any cell's AP remains unchanged in the process of interaction with other cells. Including of certain restitution rules that change the duration of functional intervals as a result of cell interaction will make it possible to advance in the study of the stability of the cell's own generation in a homogeneous tissue, multiplex reflection at the boundary, and an adequate description of such phenomena as afterdepolarization.

Our preliminary studies show that the basic properties of block and reflection do not fundamentally change as a result of such an extension of the model. However, to expand the model, quantitative representations of restitution are required that do not contradict the experimental data. In addition to the direct measurements, such data can be the results of matching the patterns of the natural cardiac rhythm and the rhythm generated by the presented model.

# 6 Conclusion

In the axiomatic 1-D model, the evolution of excitable tissue action potential is presented as a sequence of four functional states of the cell. The parameters of the model have the meaning of the duration of the intervals of the functional states of the cell. The rules of cell interaction are formulated in terms of the ratio of the durations of their functional intervals. Characterization the evolution of AP in the form of the functional intervals durations makes it possible to distinguish 4 modes of cell functioning, including the propagation and generation of oscillations in homogeneous excitable tissue.

In this paper, only one mode of the solitary waves propagation is considered, which is not accompanied by self-generation of oscillations. The conditions for the block and reflection occurrence are determinate for the incident solitary wave on the boundary of homogeneous tissues. Block and reflection have been proven to be unidirectional effects. Ultimately, the properties of the block and reflection, as unidirectional effects, are formulated at a qualitative level that does not explicitly depend on the parameters of the model.

The studied effects have a utilitarian orientation; they can serve as a basis for the construction of complex mechanisms of conduction and creation of patterns of heart rhythm, allowing improving the interpretation of cardiograms.